\pdfoutput=1

\documentclass[12pt,prd,aps,amssymb,amsmath,tightenlines,showpacs]{revtex4}
\usepackage{graphicx}

\newcommand{\cP}{\ensuremath{\mathcal{P}}}
\newcommand{\cT}{\ensuremath{\mathcal{T}}}
\newcommand{\cPT}{\ensuremath{\mathcal{PT}}}

\begin{document}

\title{$\cPT$-Symmetric Model of Immune Response}

\author{Carl M. Bender$^a$}\email{cmb@wustl.edu}
\author{Ananya Ghatak$^{a,b}$}\email{gananya04@gmail.com}
\author{Mariagiovanna Gianfreda$^{a,c}$}\email{Maria.Gianfreda@le.infn.it}

\affiliation{$^a$ Department of Physics, Washington University, St. Louis,
MO 63130, USA}
\affiliation{$^b$ Department of Physics, Indian Institute of Science,
Bangalore 560 012, India}
\affiliation{$^c$Institute of Industrial Science, University of Tokyo,
Komaba, Meguro, Tokyo 153-8505, Japan}

\date{\today}

\begin{abstract}
The study of $\cPT$-symmetric physical systems began in 1998 as a complex
generalization of conventional quantum mechanics, but beginning in 2007
experiments began to be published in which the predicted $\cPT$ phase transition
was clearly observed in classical rather than in quantum-mechanical systems.
This paper examines the $\cPT$ phase transition in mathematical models of
antigen-antibody systems. A surprising conclusion that can be drawn from these
models is that a possible way to treat a serious disease in which the antigen
concentration is growing out of bounds (and the host will die) is to inject a
small dose of a {\it second} (different) antigen. In this case there are two
possible favorable outcomes. In the unbroken-$\cPT$-symmetric phase the disease
becomes chronic and is no longer lethal while in the appropriate
broken-$\cPT$-symmetric phase the concentration of lethal antigen goes to zero
and the disease is completely cured.
\end{abstract}


\maketitle

\section{Introduction}
\label{s1}
There have been many studies of dynamical predator-prey systems that simulate
biological processes. Particularly interesting early work was done by Bell
\cite{R1}, who showed that the immune response can be modeled quite effectively
by such systems. In Bell's work the time evolution of competing concentrations
of one antigen and one antibody is studied.

The current paper shows what happens if we combine two antibody-antigen
subsystems {\it in a $\cPT$-symmetric fashion} to make an immune system in which
there are {\it two} antibodies and {\it two antigens}. An unexpected conclusion
is that even if one antigen is lethal (because the antigen concentration grows
out of bounds), the introduction of a {\it second} antigen can stabilize the
concentrations of {\it both} antigens, and thus save the life of the host. 
Introducing a second antigen may actually drive the concentration of the
lethal antigen to zero.

We say that a classical dynamical system is $\cPT$ {\it symmetric} if the
equations describing the system remain invariant under combined space reflection
$\cP$ and time reversal $\cT$ \cite{R2}. Classical $\cPT$-symmetric systems have
a typical generic structure; they consist of two coupled identical subsystems,
one having gain and the other having loss. Such systems are $\cPT$ symmetric
because under space reflection the systems with loss and with gain are
interchanged while under time reversal loss and gain are again interchanged.

Systems having $\cPT$ symmetry typically exhibit two different characteristic
behaviors. If the two subsystems are coupled sufficiently strongly, then the
gain in one subsystem can be balanced by the loss in the other and thus the
total system can be in equilibrium. In this case the system is said to be in an 
{\it unbroken} $\cPT$-symmetric phase. (One visible indication that a linear
system is in an unbroken phase is that it exhibits Rabi oscillations in which
energy oscillates between the two subsystems.) However, if the subsystems are
weakly coupled, the amplitude in the subsystem with gain grows while the
amplitude in the subsystem with loss decays. Such a system is not in equilibrium
and is said to be in a {\it broken} $\cPT$-symmetric phase. Interestingly, if
the subsystems are very strongly coupled, it may also be in a broken
$\cPT$-symmetric phase because one subsystem tends to drag the other subsystem.

A simple linear $\cPT$-symmetric system that exhibits a $\cPT$ phase transition
from weak to moderate coupling and a second transition from moderate to strong
coupling consists of a pair of coupled oscillators, one with damping and the
other with antidamping. Such a system is described by the pair of linear
differential equations
\begin{equation}
\ddot{x}+\dot{x}+\omega^2 x=\epsilon xy,\qquad 
\ddot{y}-\dot{y}+\omega^2 y=\epsilon xy.
\label{e1}
\end{equation}
This system is invariant under combined parity reflection $\cP$, which
interchanges $x$ and $y$, and time reversal $\cT$, which replaces $t$ with $-t$.
Theoretical and experimental studies of such a system may be found in
Refs.~\cite{R3,R4}. For an investigation of a $\cPT$-symmetric system of many
coupled oscillators see Ref.~\cite{RA}. Experimental studies of $\cPT$-symmetric
systems may be found in Refs.~\cite{S1,S2,S3,S4,S5,S6,S7,S8,S9,S10}.

It is equally easy to find physical nonlinear $\cPT$-symmetric physical systems.
For example, consider a solution containing the oxidizing reagent potassium
permanganate ${\rm KMnO}_4$ and a reducing agent such as oxalic acid ${\rm
COOH}_2$. The reaction of these reagents is self-catalyzing because the
presence of manganous ${\rm Mn}^{+2}$ ions increases the speed of the reaction.
The chemical reaction in the presence of oxalic acid is
$${\rm MnO}_4^{-1}+{\rm Mn}^{+2}\longrightarrow 2{\rm Mn}^{+2}.$$
Thus, if $x(t)$ is the concentration of permanganate ions and $y(t)$ is the
concentration of manganous ions, then the rate equation is
\begin{equation}
\dot{x}=-kxy,\qquad \dot{y}=kxy,
\label{e2}
\end{equation}
where $k$ is the rate constant. This system is $\cPT$ invariant, where $\cP$
exchanges $x$ and $y$ and $\cT$ replaces $t$ with $-t$. For this system, the
$\cPT$ symmetry is always broken; the system is not in equilibrium.

The Volterra (predator-prey) equations are a slightly more complicated
$\cPT$-symmetric nonlinear system:
\begin{equation}
\dot{x}=a x-bx y,\qquad \dot{y}=-ay+bxy.
\label{e3}
\end{equation}
This system is oscillatory and thus we say that the $\cPT$ symmetry is unbroken.
These equations are discussed in Ref.~\cite{R5}. A nonlinear $\cPT$-symmetric
system of equations that exhibits a phase transition between broken and unbroken
regions may be found in Ref.~\cite{R5.5}.

In analyzing elementary systems like that in (\ref{e1}), which are described by
constant-coefficient differential equations, the usual procedure is to make the
{\it ansatz} $x(t)=Ae^{i\nu t}$ and $y(t)=Be^{i\nu t}$. This reduces the system
of differential equations to a polynomial equation for the frequency $\nu$. We
then associate unbroken (or broken) phases with real (or complex) frequencies
$\nu$. If $\nu$ is real, the solutions to both equations are oscillatory and
remain bounded, and this indicates that the physical system is in dynamic
equilibrium. However, if $\nu$ is complex, the solutions grow or decay
exponentially with $t$, which indicates that the system is not in equilibrium.

For more complicated nonlinear $\cPT$-symmetric dynamical systems, we still say
that the system is in a phase of broken $\cPT$ symmetry if the solutions grow or
decay with time or approach a limit as $t\to\infty$ because the system is not in
dynamic equilibrium. In contrast, if the variables oscillate and remain bounded
as $t$ increases we say that the system is in a phase of unbroken $\cPT$
symmetry. However, in this case the time dependence of the variables is
unlikely to be periodic; such systems usually exhibit {\it almost periodic} or
{\it chaotic} behavior.

\begin{figure}[h!]
\begin{center}
\includegraphics[scale=0.36]{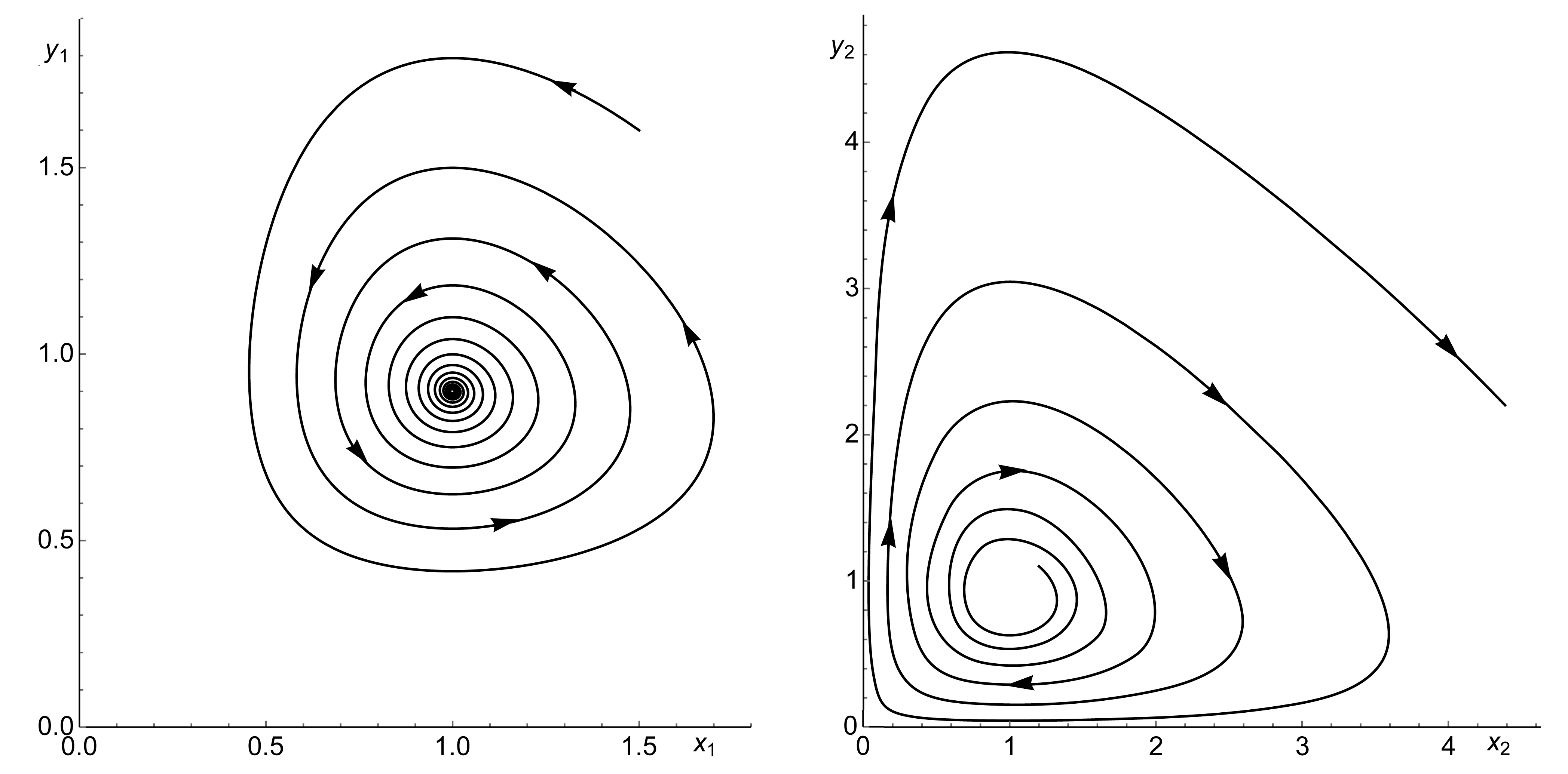}
\end{center}
\caption{Left panel: An inspiral trajectory plotted in the $(x_1,y_1)$ plane for
the dynamical subsystem (\ref{e4}) with $c=0.1$. The initial conditions are $x_1
(0)=1.5,~y_1(0)=1.6$. Right panel: An outspiral trajectory for (\ref{e5}) in the
$(x_2,y_2)$ plane with $c=0.1$. The initial conditions are $x_2(0)=1.2,~y_2(0)=
1.1$. In the left panel $t$ ranges from $0$ to $100$ and in the right panel $t$
ranges from $0$ to $45$.}
\label{f1}
\end{figure}

To illustrate these possibilities we construct a more elaborate $\cPT$-symmetric
system of nonlinear equations by combining a two-dimensional dynamical subsystem
whose trajectories are {\it inspirals} with another two-dimensional dynamical
subsystem whose trajectories are {\it outspirals}. For example, consider the
subsystem
\begin{eqnarray}
\dot{x}_1&=&x_1-x_1y_1-cx_1^2,\nonumber\\
\dot{y}_1&=&-y_1+x_1y_1.
\label{e4}
\end{eqnarray}
This system has two saddle points and one stable spiral point, as shown in
Fig.~\ref{f1} (left panel).

\begin{figure}[h!]
\begin{center}
\includegraphics[scale=0.36]{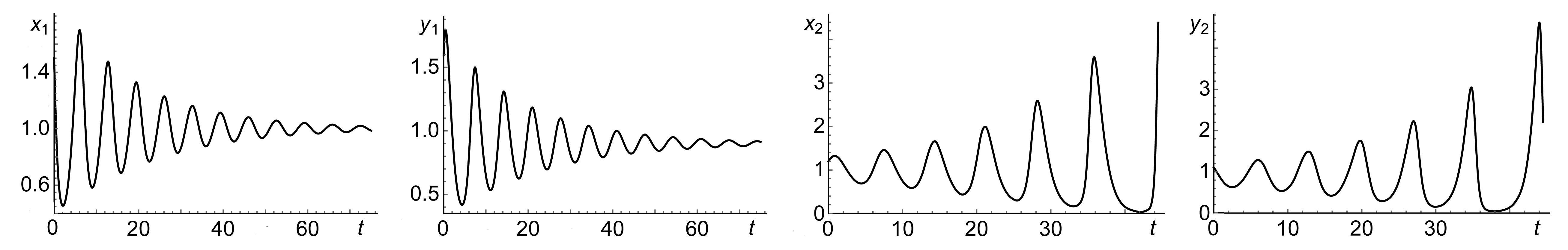}
\end{center}
\caption{Inspiral and outspiral for the initial conditions in Fig.~\ref{f1}.
The four variables $x_1(t)$, $y_1(t)$, $x_2(t)$, and $y_2(t)$ are plotted as
functions of $t$.}
\label{f2}
\end{figure}

Next, we consider the $\cPT$ reflection ($x_1\to x_2$, $y_1\to y_2$, $t\to-t$)
of the subsystem in (\ref{e4}):
\begin{eqnarray}
\dot{x}_2&=&-x_2+x_2 y_2+cx_2^2,\nonumber\\
\dot{y}_2&=&y_2-x_2 y_2.
\label{e5}
\end{eqnarray}
The trajectories of this system are outspirals, as shown in Fig.~\ref{f1} (right
panel). The time evolution of the four dynamical variables in Fig.~\ref{f1},
$x_1(t)$ and $y_1(t)$, $x_2(t)$ and $y_2(t)$, is shown in Fig.~\ref{f2}.

\begin{figure}[h!]
\begin{center}
\includegraphics[scale=0.36]{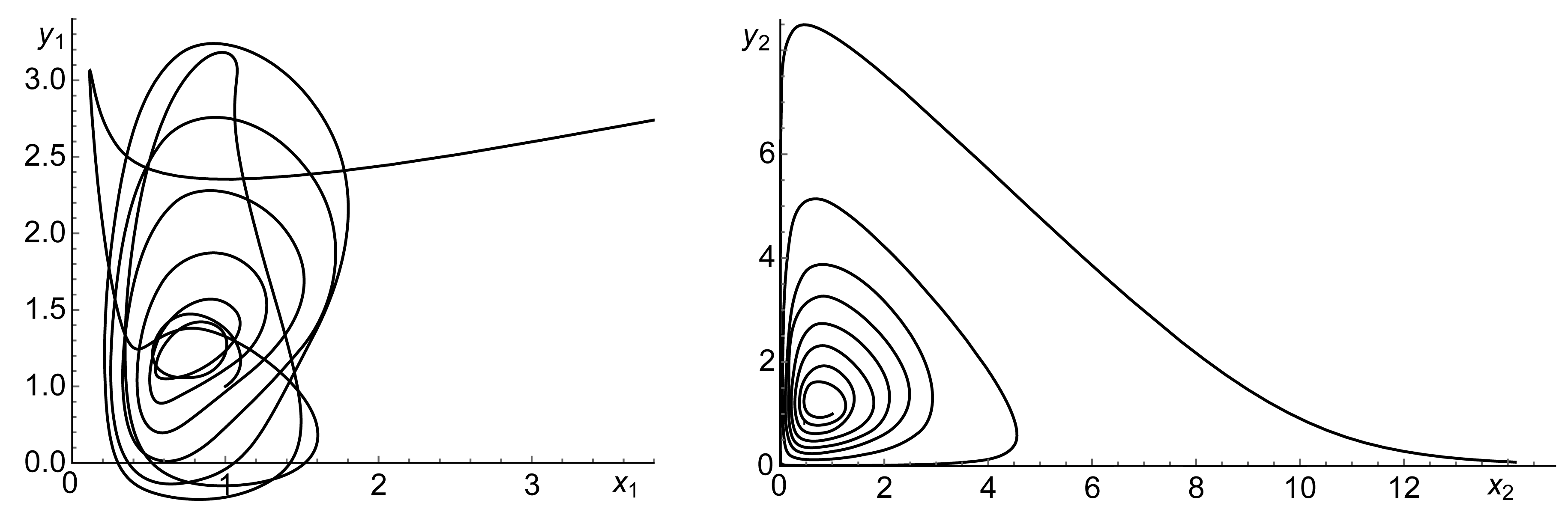}
\end{center}
\caption{$\cPT$-symmetric system (\ref{e6}) in a broken-$\cPT$-symmetric phase,
as indicated by the outspiral behavior in the $[x_1(t),y_1(t)]$ and $[x_2(t),
y_2(t)]$ planes. The parametric values are $c=0.2$, $f=0.2$, and $g=0.5$ and the
initial conditions are $x_1(0)=y_1(0)=x_2(0)=y_2(0)=1.0$. In these plots $t$
ranges from $0$ to $60$.}
\label{f3}
\end{figure}

\begin{figure}[h!]
\begin{center}
\includegraphics[scale=0.36]{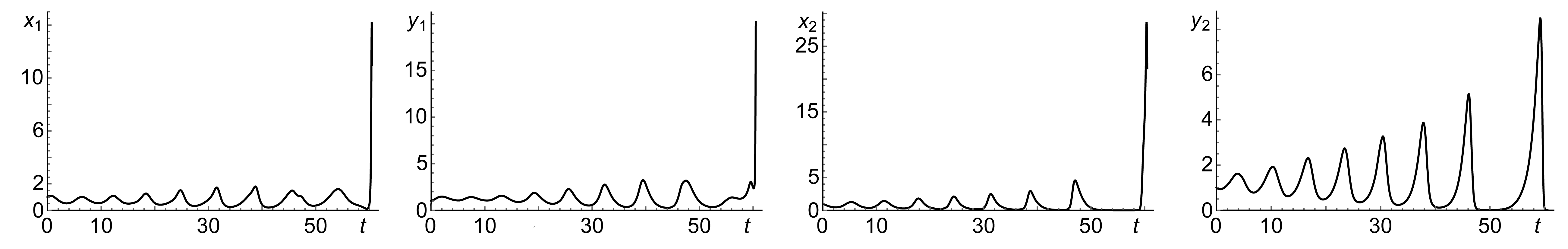}
\end{center}
\caption{Time dependence of $x_1(t)$, $y_1(t)$, $x_2(t)$, $y_2(t)$ for the
parametric values and initial conditions shown in Fig.~\ref{f3}.}
\label{f4}
\end{figure}

Let us now couple the two subsystems in (\ref{e4}) and (\ref{e5}) in such a way
that the $\cPT$ symmetry is preserved. The resulting dynamical system obeys
the nonlinear equations
\begin{eqnarray}
\dot{x}_1&=&x_1-x_1y_1-cx_1^2+gx_1x_2,\nonumber\\
\dot{y}_1&=&-y_1+x_1y_1+fy_1y_2,\nonumber\\ 
\dot{x}_2&=&-x_2+x_2y_2+cx_2^2-gx_1x_2,\nonumber\\
\dot{y}_2&=&y_2-x_2y_2-fy_1y_2
\label{e6}
\end{eqnarray}
in which $f$ and $g$ are the coupling parameters. This system has a wide range
of possible behaviors. For example, for the parametric values $c=0.2$, $f=0.2$,
and $g=0.5$ and the initial conditions $x_1(0)=y_1(0)=x_2(0)=y_2(0)=1.0$ we can
see from Figs.~\ref{f3} and \ref{f4} that the system is in a
broken-$\cPT$-symmetric phase.

\begin{figure}[h!]
\begin{center}
\includegraphics[scale=0.44]{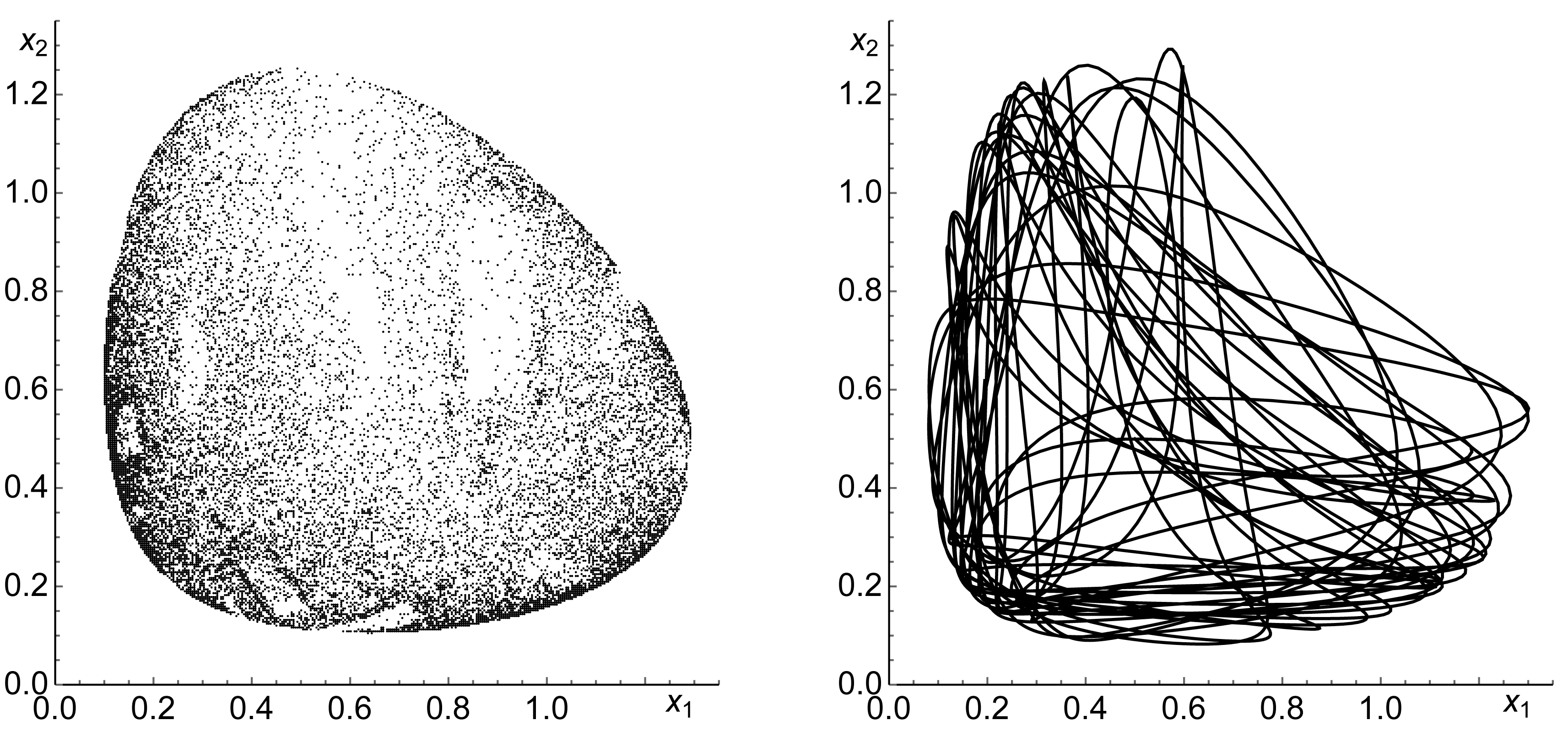}
\end{center}
\caption{System (\ref{e6}) in a phase of chaotic unbroken $\cPT$ symmetry. The
parametric values are $c=0.2$, $f=0.5$, $g=0.3$ and the initial conditions are
$x_1(0)=y_1(0)=x_2(0)=y_2(0)=0.5$. Left panel: Poincar\'e plot of $x_1$ versus
$x_2$ when $y_2=0.75$. The two-dimensional scatter of dots indicates that the
system is chaotic. In this plot $t$ ranges from $0$ to $100,000$. Right panel:
A plot of $x_1(t)$ versus $x_2(t)$ for $t$ ranging from $0$ to $300$.}
\label{f5}
\end{figure}

\begin{figure}[h!]
\begin{center}
\includegraphics[scale=0.44]{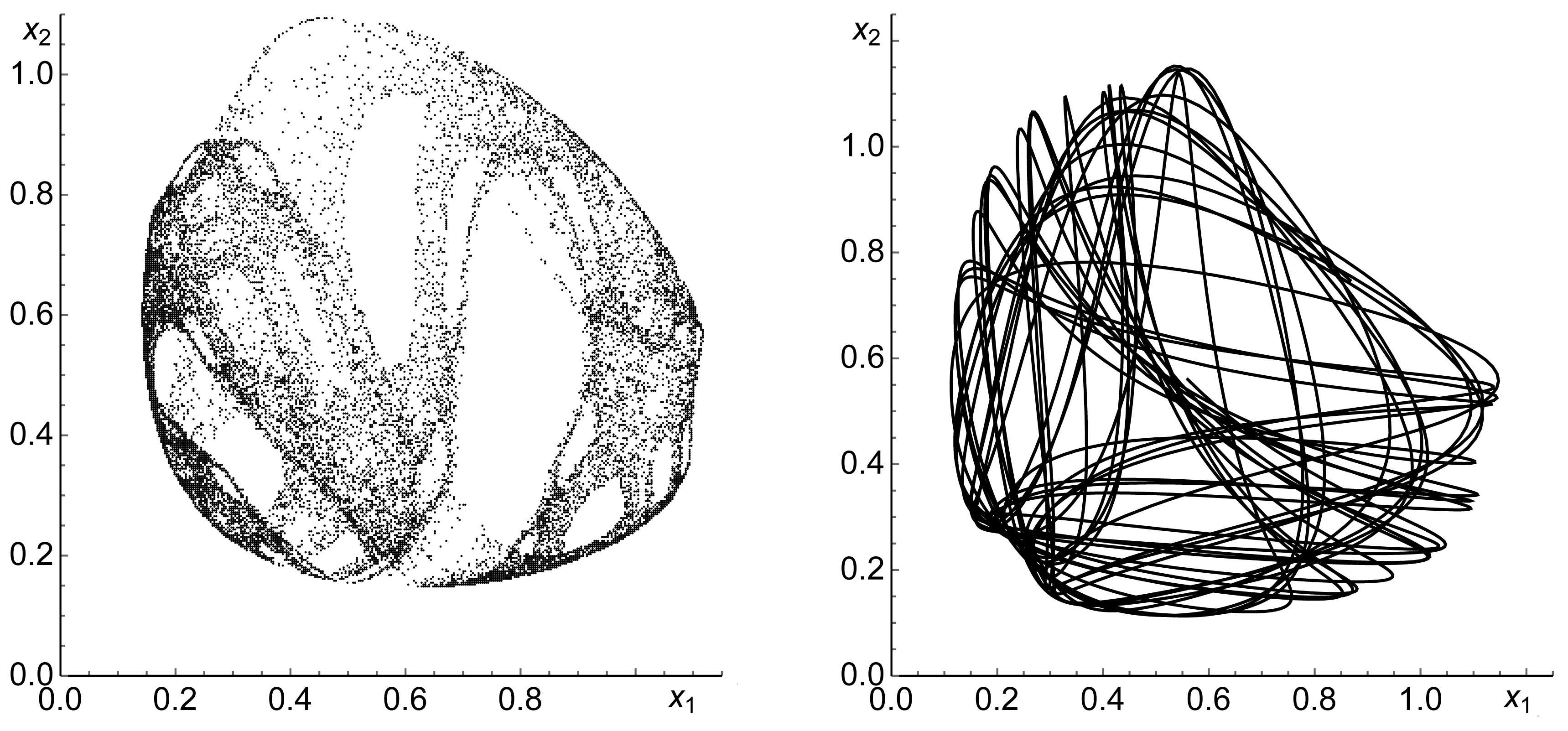}
\end{center}
\caption{System in Fig.~ref{f5} in a different chaotic state of unbroken $\cPT$
symmetry. The parametric values and the ranges of $t$ are the same as in
Fig.~\ref{f5}, but the initial conditions are now $x_1(0)=y_1(0)=x_2(0)=y_2(0)=
0.56$.}
\label{f6}
\end{figure}

\begin{figure}[h!]
\begin{center}
\includegraphics[scale=0.45]{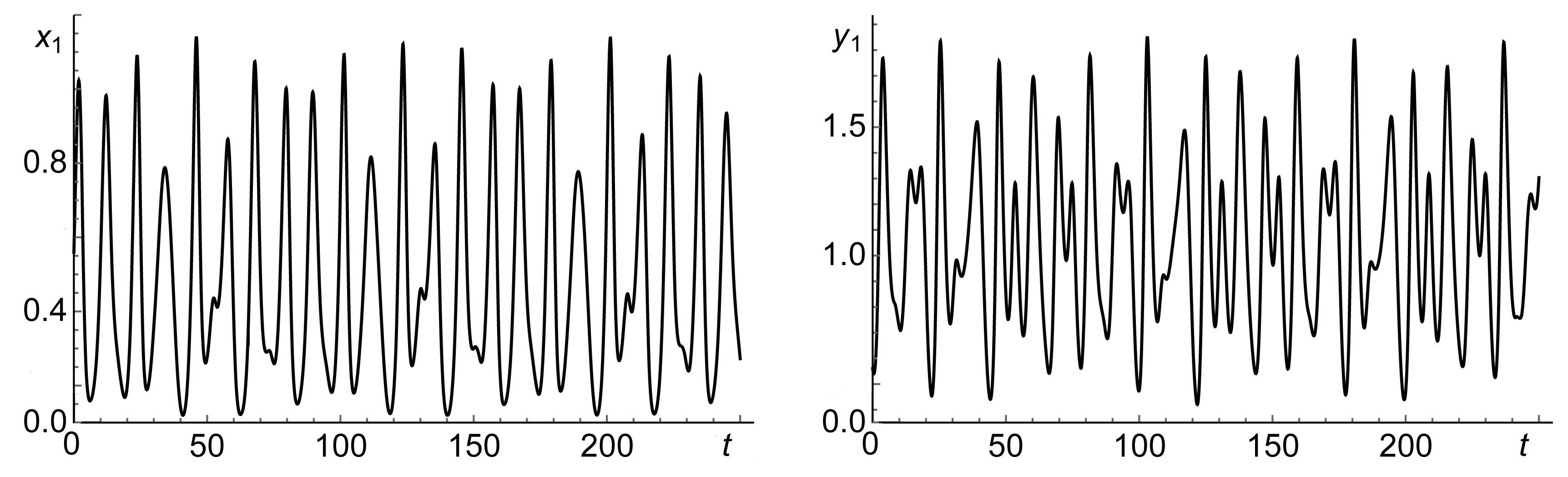}
\vspace{.3mm}\\
\includegraphics[scale=0.45]{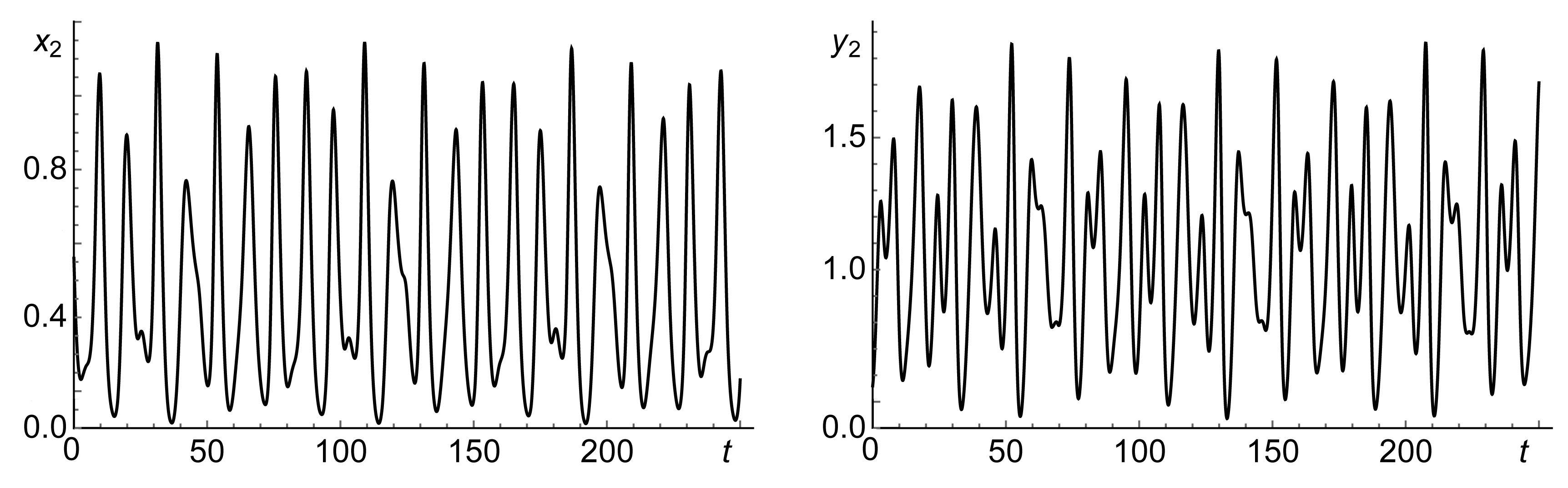}
\end{center}
\caption{The system in Fig.~\ref{f6} plotted as a function of time. The chaotic
behavior can be seen as the uneven oscillations. These oscillations are
reminiscent of a trajectory under the influence of a pair of strange
attractors.}
\label{f78}
\end{figure}

\begin{figure}[h!]
\begin{center}
\includegraphics[scale=0.44]{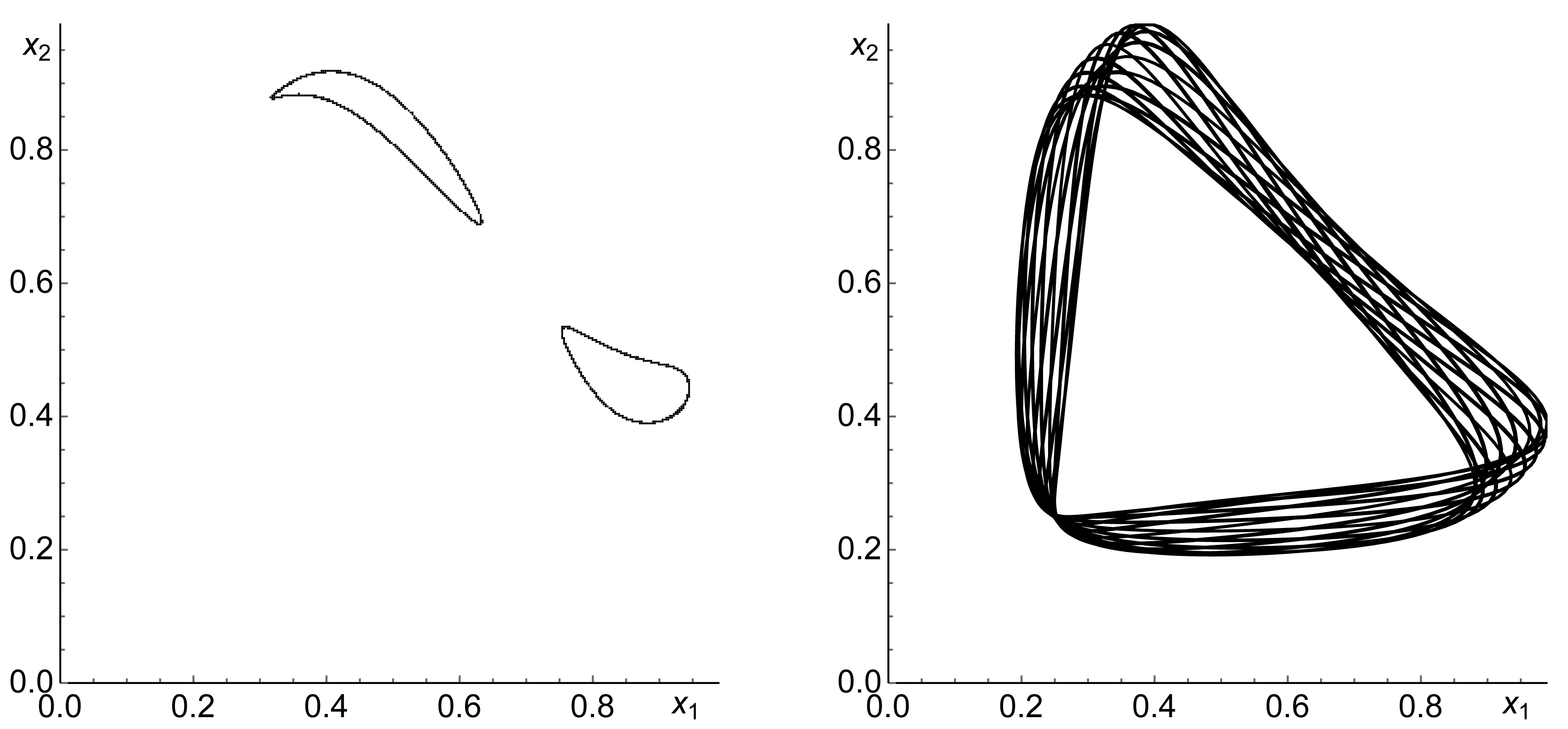}
\end{center}
\caption{System (\ref{e6}) in a state of unbroken $\cPT$ symmetry. The
parametric values and the time ranges are the same as in Fig.~\ref{f5}, but the
initial conditions are $x_1(0)=y_1(0)=x_2(0)=y_2(0)=0.68$. The presence of
one-dimensional {\it islands} in the Poincar\'e plot (left panel) shows that the
time evolution of the system is almost periodic.}
\label{f9}
\end{figure}

\begin{figure}[h!]
\begin{center}
\includegraphics[scale=0.44]{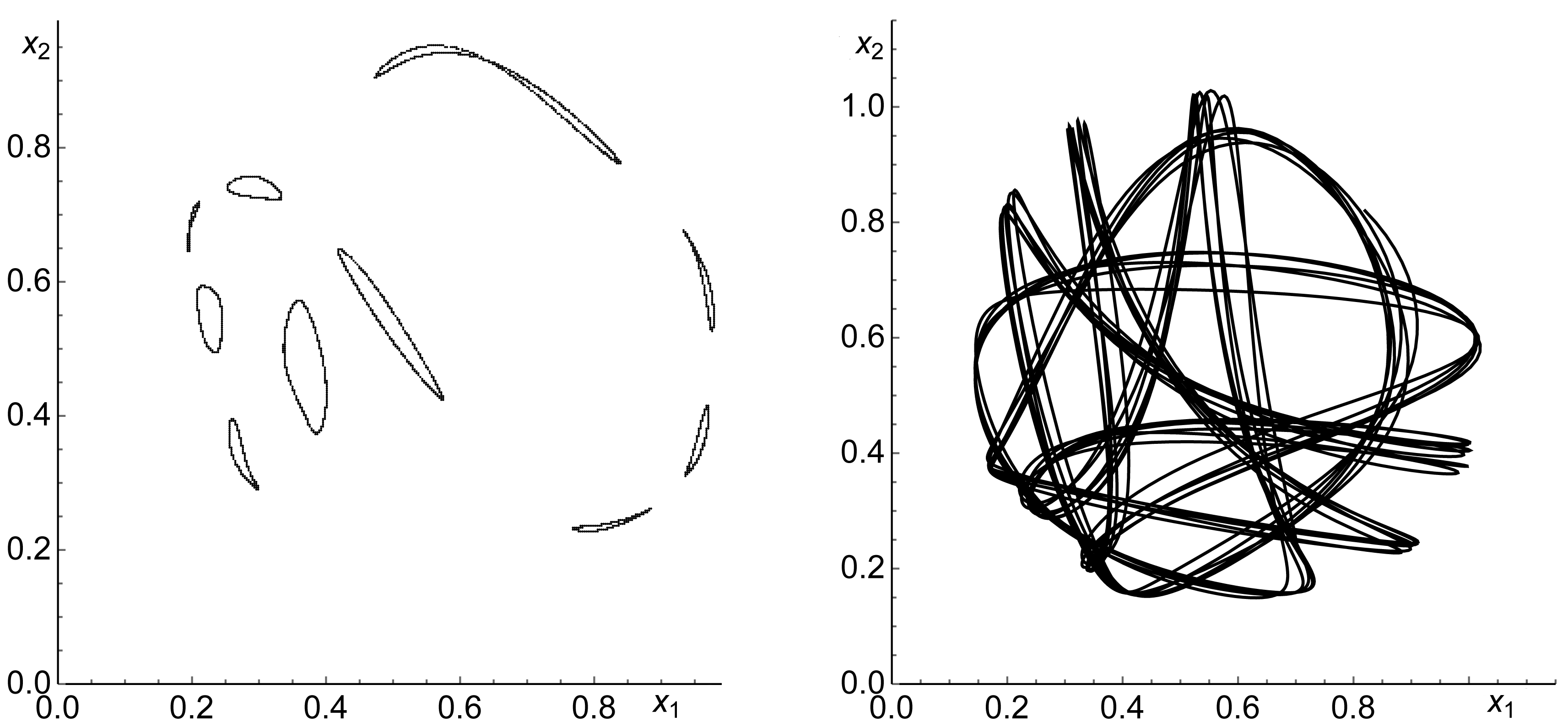}
\end{center}
\caption{System (\ref{e6}) in a different state of unbroken $\cPT$ symmetry. The
parametric values and the ranges of $t$ are the same as in Fig.~\ref{f9}, but
the initial conditions are $x_1(0)=y_1(0)=x_2(0)=y_2(0)=0.82$. The Poincar\'e
plot (left panel) again shows that the time evolution of the system is almost
periodic.}
\label{f10}
\end{figure}

\begin{figure}[h!]
\begin{center}
\includegraphics[scale=0.45]{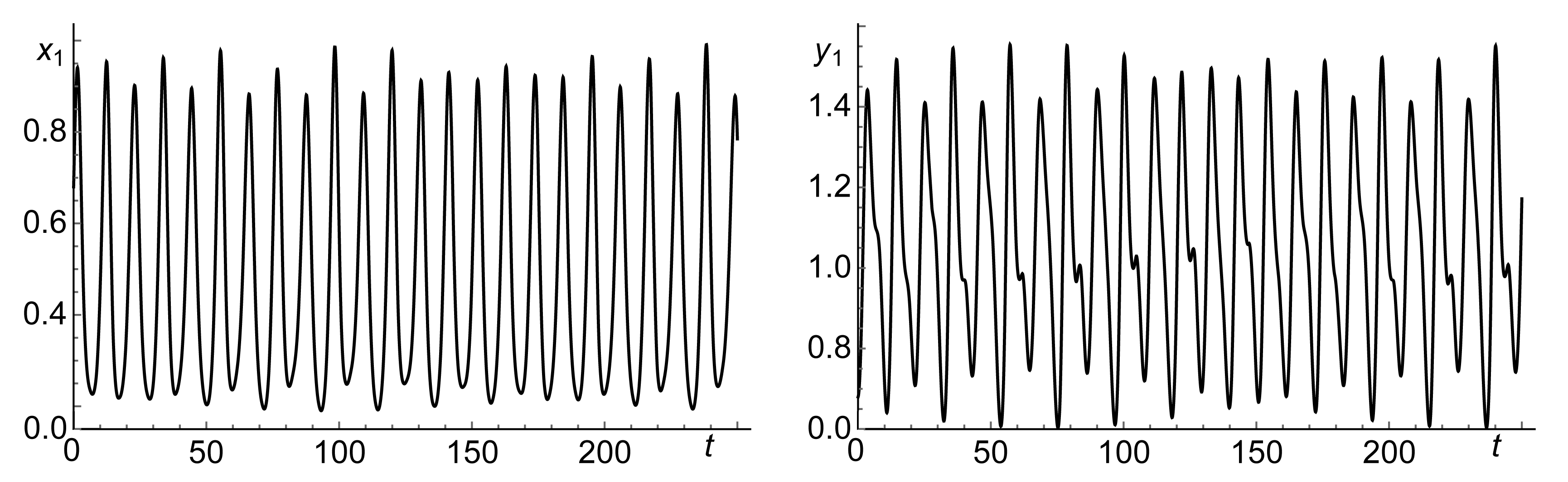}
\vspace{.3mm}\\
\includegraphics[scale=0.45]{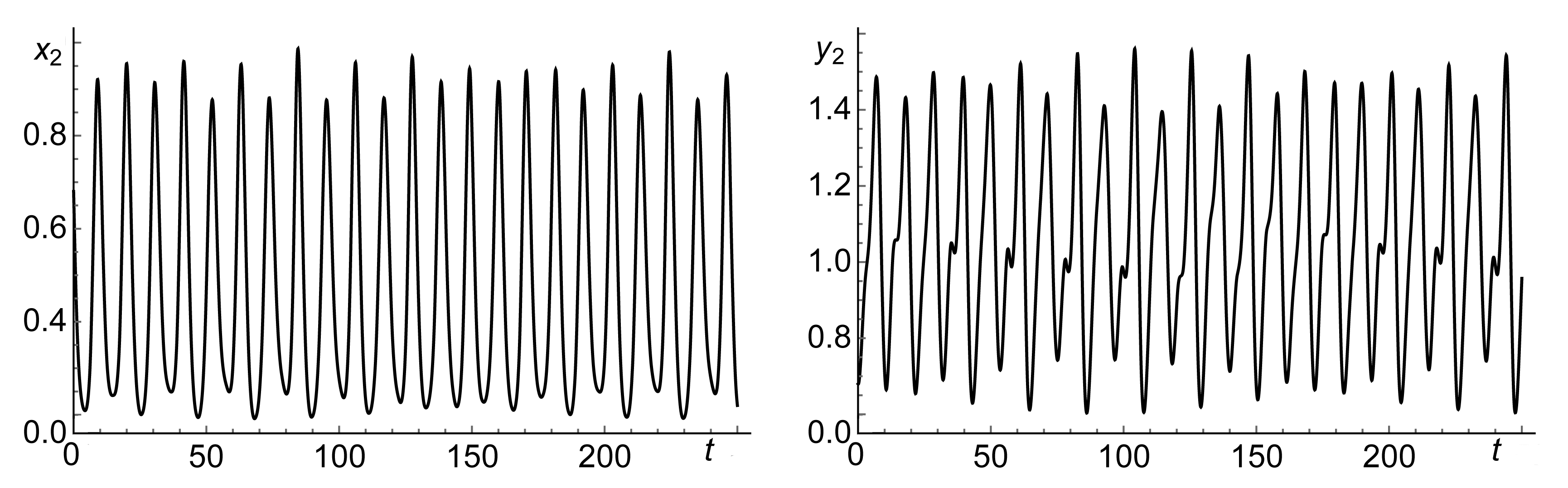}
\end{center}
\caption{The system in Fig.~\ref{f10} plotted as a function of time. The almost
periodic behavior is particularly evident in the graphs on the left, where the
oscillations are quite regular.}
\label{f1112}
\end{figure}

When the coupling parameters are chosen so that the system (\ref{e6}) is in a
phase of unbroken $\cPT$ symmetry, the initial conditions determine whether the
behavior is chaotic or almost periodic. For example, for the same parametric
values $c=0.2$, $f=0.2$, and $g=0.3$ the system in (\ref{e6}) is in an
unbroken-$\cPT$-symmetric phase. Two qualitatively different behaviors of
unbroken $\cPT$ symmetry are illustrated in Figs.~\ref{f5}, \ref{f6}, \ref{f78}
and \ref{f9}, \ref{f10}, and \ref{f1112}. The first three figures display the
system in two states of chaotic equilibrium and the next three show the system
in two states of almost-periodic equilibrium. The Poincar\'e plots in
Figs.~\ref{f5} and \ref{f6} (left panels) and Figs.~\ref{f9} and \ref{f10}
(left panels) distinguish between chaotic and almost periodic behavior.

\begin{figure}[h!]
\begin{center}
\includegraphics[scale=0.43]{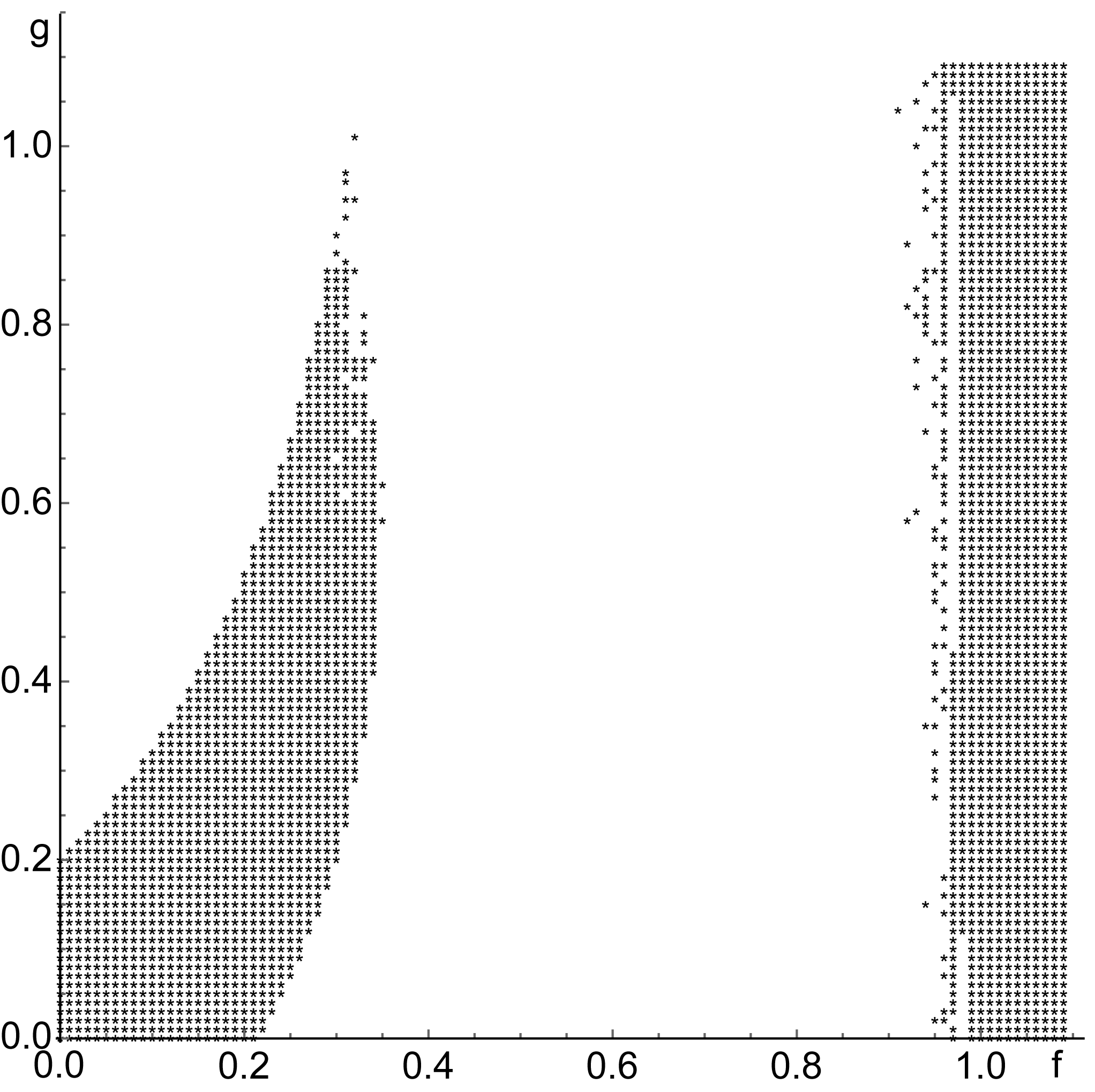}
\end{center}
\caption{A region of the $(f,g)$ plane for the system (\ref{e6}) with the 
parametric value $c=0.2$. The initial conditions are $x_1(0)=y_1(0)=x_2(0)=y_2
(0)=1$. The dots correspond to parametric values $(f,g)$ in the region of broken
$\cPT$ symmetry, and the white space corresponds to the region of unbroken
$\cPT$ symmetry. The edges of the regions are not completely sharp; it can be
difficult to determine the precise location of the boundary curves separating
broken and unbroken regions because this requires integrating for extremely long
times.}
\label{f13}
\end{figure}

The choice of coupling parameters usually (but not always) determines whether
the system is in an unbroken or a broken $\cPT$-symmetric phase. To demonstrate
this, we take $c=0.2$ and examine the time evolution for roughly 11,000 values
of the parameters $f$ and $g$. Figure~\ref{f13} indicates the values of $f$ and
$g$ for which the system is in a broken or an unbroken (chaotic or almost
periodic) phase.

Having summarized the possible behaviors of coupled $\cPT$-symmetric dynamical
subsystems, in Sec.~\ref{s2} we construct and examine in detail a
$\cPT$-symmetric dynamical model of an antigen-antibody system containing {\it
two} antigens and {\it two} antibodies. This system is similar in structure to
that in (\ref{e6}). We show that in the unbroken region the concentrations of
antigens and antibodies generally become chaotic and we interpret this as a
chronic infection. However, in the unbroken regions there are two possibilities;
either the antigen concentration grows out of bounds (the host dies) or else the
antigen concentration falls to zero (the disease is completely cured). Some
concluding remarks are given in Sec.~\ref{s3}.

\section{Dynamical Model of Competing Antibody-Antigen Systems}
\label{s2}

Infecting an animal with bacteria, foreign cells, or virus may produce an immune
response. The foreign material provoking the response is called an {\it antigen}
and the immune response is characterized by the production of {\it antibodies},
which are molecules that bind specifically to the antigen and cause its
destruction. The time-dependent immune response to a replicating antigen may be
treated as a dynamical system with interacting populations of the antigen, the
antibodies, and the cells that are involved in the production of antibodies. A
detailed description of such an immune response would be extremely complicated
so in this paper we consider a simplified mathematical model of the immune
response proposed by Bell \cite{R1}. Bell's paper introduces a simple model in
which the multiplication of antigen and antibodies is assumed to be governed by
Lokta-Volterra-type equations, where the antigen plays the role of prey and the
antibody plays the role of predator. While such a model may be an unrealistic
simulation of an actual immune response, Bell argues that this mathematical
approach gives a useful qualitative and quantitative description.

Following Bell's paper we take the variable $x_1(t)$ to represent the
concentration of antibody and the variable $y_1(t)$ to represent the
concentration of antigen at time $t$. Assuming that the system has an unlimited
capability of antibody production, Bell's dynamical model describes the time
dependence of antigen and antibody concentrations by the differential equations
\begin{eqnarray}
\dot{x}_1&=&-\lambda_2\,x_1+\alpha_2\,u(x_1,y_1),\nonumber\\
\dot{y}_1&=&\lambda_1\,y_1-\,\alpha_1\,v(x_1,y_1).
\label{e7}
\end{eqnarray}
According to (\ref{e7}), the antigen concentration $y_1$ increases at a constant
rate $\lambda_1$ if the antibody $x_1$ is are not present. As soon as antigens
are bound to antibodies, the antibodies start being eliminated at the constant
rate $\alpha_1$. Analogously, the concentration of antibody $x_1$ decays with
constant rate $\lambda_2$ in the absence of antigens, while binding of antigens
to antibodies stimulates the production of antibody $x_1$ with constant rate
$\alpha_2$. The functions $u(x_1,y_1)$ and $v(x_1,y_1)$ denote the
concentrations of bound antibodies and bound antigens. Assuming that $u(x_1,y_1)
=v(x_1,y_1)$, an approximate expression for the concentration of bound antigens
and antibodies is
\begin{equation}
u(x_1,y_1)=v(x_1,y_1)=\frac{k\,x_1\,y_1}{1+k(x_1+y_1)}\equiv F(x_1,y_1),
\label{e8}
\end{equation}
where $k$ is called an {\it association constant}. With the scalings $kx_1\to
x_1$ and $ky_1\to y_1$ and the change of variable
$$s=\int_0^t\,dt^\prime\,\left[1+x_1(t^\prime)+y_1(t^\prime)\right]^{-1},$$
system (\ref{e7}) becomes
\begin{eqnarray}
\frac{dx_1}{ds}&=&-\lambda_2\,x-\lambda_2\,x_1^2+(\alpha_2-\lambda_2)x_1y_1,
\nonumber\\
\frac{dy_1}{ds}&=&\lambda_1\,y_1+\lambda_1\,y_1^2-(\alpha_1-\lambda_1)x_1y_1.
\label{e9}
\end{eqnarray}

The system (\ref{e9}) exhibits four different behaviors:
\begin{itemize}
\item[(1)] If
$R\equiv\alpha_1\alpha_2-\alpha_1\lambda_2-\alpha_2\lambda_1<0$, there is
unbounded monotonic growth of antigen.
\item[(2)] If $R>0$ and $\alpha_1>\alpha_2$, there is an outspiral (oscillating
growth of antigen).
\item[(3)] If $R>0$ and $\alpha_1<\alpha_2$, there is an inspiral (the
antigen approaches a limiting value in an oscillatory fashion).
\item[(4)] If $R>0$ and $\alpha_1=\alpha_2$, the system exhibits exactly
periodic oscillations. This behavior is unusual in a nonlinear system and
indeed (\ref{e6}) does not exhibit exact periodic behavior.
\end{itemize}

\subsection{$\cPT$-symmetric interacting model}
Subsequent to Bell's paper \cite{R1} there have been many studies that use 
two-dimensional dynamical models to examine the antigen-antibody interaction
\cite{RAA}.
However, in this paper we construct a {\it four}-dimensional model consisting of
two antigens and two antibodies. Let us assume that an antigen $y_1$ attacks an
organism and that the immune response consists of creating antibodies $x_1$ as
described by (\ref{e7}). However, we suppose that the organism has a second
system of antibodies and antigens $(x_2,y_2)$. This second subsystem plays the
role of a $\cPT$-symmetric partner of the system $(x_1,y_1)$, where parity $\cP$
interchanges the antibody $x_1$ with the antigen $y_2$ and the antigen $y_1$
with the antibody $x_1$,
$$\cP:\,x_1\to y_2,\quad x_2\to y_1,$$ 
and time reversal $\cT$ makes the replacement $t\to -t$. The time evolution of
this new antibody-antigen system is regulated by the equations
\begin{eqnarray}
\dot{x}_2&=&-\lambda_1\,x_2+\alpha_1\,F(x_2,y_2),\nonumber\\
\dot{y}_2&=&\lambda_2\,y_2-\,\alpha_2\,F(x_2,y_2).
\label{e10}
\end{eqnarray}
We assume that the interaction between antibody $x_2$ and antigen $y_2$ is
controlled by the same constant $k$ as in (\ref{e8}).

We assume that because antibodies may have many possible binding sites, $x_1$
can also bind to antigen $y_2$ and that antibody $x_2$ can also bind to antigen
$y_1$. Moreover, for this model we assume that we can scale the dynamical
variables so that this interaction is the same as the interaction $x_1-y_1$ and
$x_2-y_2$. This means that after the scaling $kx_1\to x_1$, $ky_1\to y_1$,
$kx_2\to x_2$, and $ky_2\to y_2$, the dynamical behavior of the total system
$(x_1,y_1,x_2,y_2)$ is described by
\begin{eqnarray}
\dot{x}_1&=&-\lambda_2x_1+\alpha_2\frac{x_1y_1}{1+x_1+y_1}+g\frac{x_1y_2}{1+x_1+
y_2},\nonumber\\
\dot{y}_1&=&\lambda_1y_1-\alpha_1\frac{x_1y_1}{1+x_1+y_1}-f\frac{x_2y_1}{1+x_2+
y_1},\nonumber\\
\dot{x}_2&=&-\lambda_1x_2+\alpha_1\frac{x_2y_2}{1+x_2+y_2}+f\frac{x_2y_1}{1+x_2+
y_1},\nonumber\\
\dot{y}_2&=&\lambda_2y_2-\alpha_2\frac{x_2y_2}{1+x_2+y_2}-g\frac{x_1y_2}{1+x_1+
y_2}.
\label{e11}
\end{eqnarray}
The production of the antibody $x_2$ is stimulated by the presence of the
antigen $y_2$. The terms involving the parameter $f$ describe the production of
additional antibodies $x_2$ and additional elimination of antigens $y_1$.
Similarly, $g$ terms describe the production of new antibodies $x_1$ and
additional elimination of antigens $y_2$.

\subsection{Hamiltonian for (\ref{e11})}
We remark that the system (\ref{e11}) with $\alpha_1=\alpha_2=\alpha$ can be
derived from the Hamiltonian \cite{R6}
\begin{equation}
H=\alpha x^{-\lambda_1/\alpha}y^{-\lambda_2/\alpha}(1+x+y).
\label{e12}
\end{equation}
That is, it can be recovered from the Hamilton equations
\begin{equation}
\frac{\partial H}{\partial x}=J_{12}(x,y)\,\dot{y},\qquad
\frac{\partial H}{\partial y}=J_{21}(x,y)\,\dot{x},
\label{e13}
\end{equation}
where
$$J_{21}(x,y)=-J_{12}(x,y)=x^{-1-\lambda_1/\alpha}\,y^{-1-\lambda_2/\alpha}.$$

\subsection{Numerical results}

\begin{figure}[h!]
\begin{center}
\includegraphics[scale=0.65]{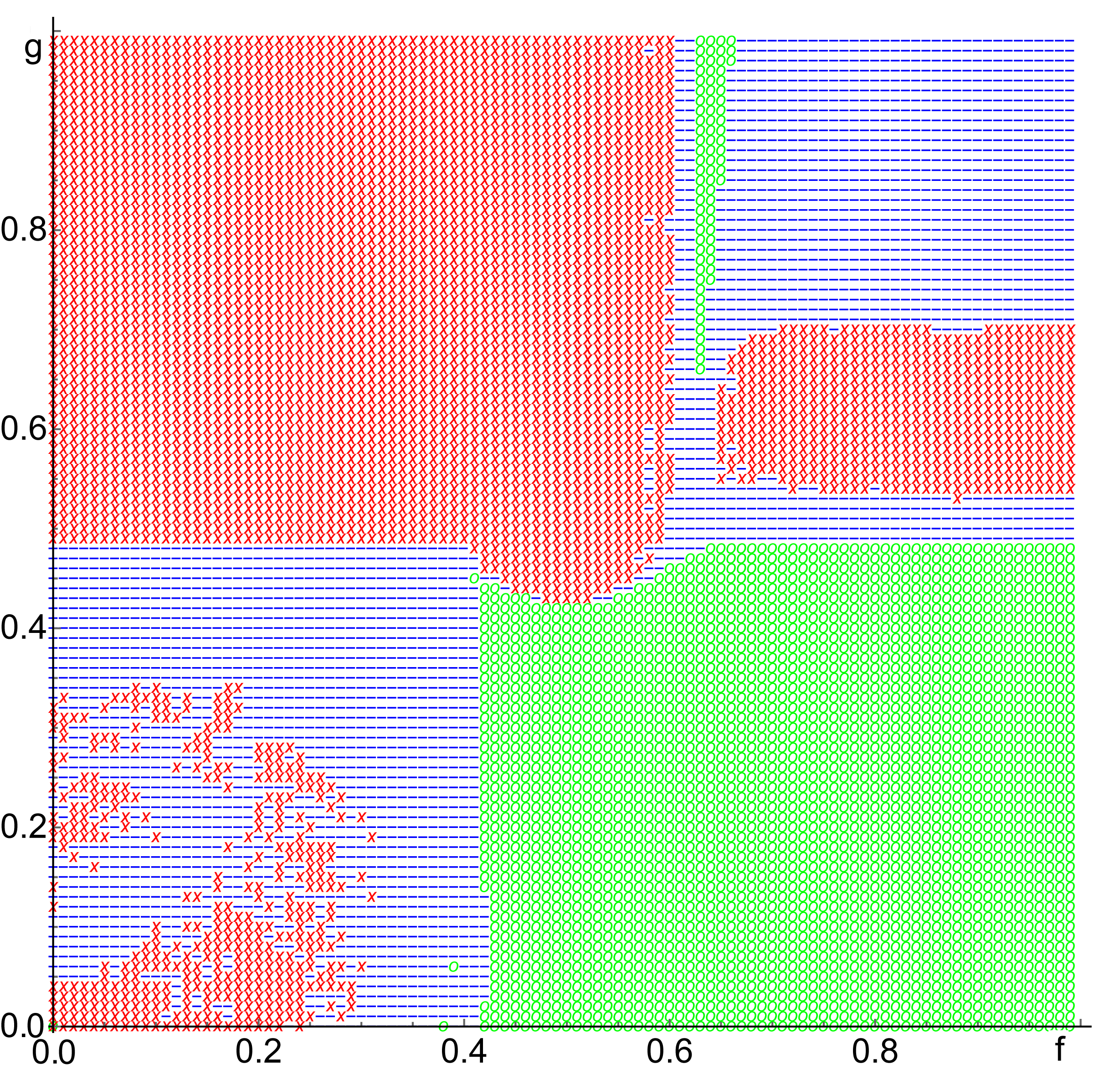}
\end{center}
\caption{[Color online] Portion of the $(f,g)$ coupling-parameter plane for the
$\cPT$-symmetric immune-response system (\ref{e11}) showing regions of broken
and unbroken $\cPT$ symmetry. We take as initial conditions $x_1(0)=y_1(0)=1$ 
and $x_2(0)=x_2(0)=0.01$; that is, we assume that the disease associated with
antigen-antibody 1 is well established and that at $t=0$ a very small amount of
antigen-antibody 2 is injected. Points in the unbroken region are indicated as
hyphens (blue). In this region the concentrations $x_1,y_1,x_2,y_2$ are all
oscillatory in time. In general, depending on the initial conditions, the
solutions can be either almost periodic or chaotic. However, as shown in
Fig.~\ref{f18}, the solutions to (\ref{e11}) are chaotic. Thus, in this region
the introduction of antigen-antibody 2 makes the potentially lethal infection
chronic. The regions whose points are indicated as o's (green) and x's (red)
have broken $\cPT$ symmetry. In the x regions the solutions oscillate and grow
out of bounds. In the o regions $x_1(t)$ and $y_1(t)$ vanish and $x_2(t)$ and
$y_2(t)$ approach small finite values as $t\to\infty$. Thus, in the x regions
the host dies, but in the o regions the disease due to antigen $y_1$ is
completely cured.}
\label{f14}
\end{figure}

Figure~\ref{f14} displays a phase diagram of the $\cPT$-symmetric model in 
(\ref{e11}), where we have taken $\lambda_1=\lambda_2=0.1$, $\alpha_1=0.6$, and
$\alpha_2=0.5$. In this figure a portion of the $(f,g)$ plane is shown and the
regions of broken and unbroken $\cPT$ symmetry are indicated.
Unbroken-$\cPT$-symmetric regions are indicated as hyphens (blue online). There
are two kinds of broken-$\cPT$-symmetric regions; x's (red online) indicate
solutions that grow out of bounds and o's (green online) indicate solutions for
which the concentration of antigen $y_1$ approaches 0.

\begin{figure}[h!]
\begin{center}
\includegraphics[scale=0.38]{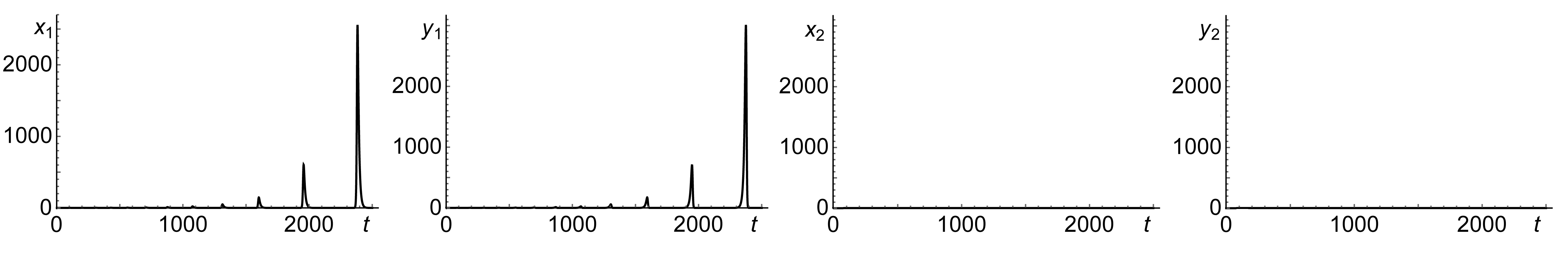}
\end{center}
\caption{An organism that does not survive an antigen attack. Here, the
antigen-antibody dynamics in (\ref{e11}) is described by (\ref{e7}) because we
take $x_2(0)=y_2(0)=0$ and thus $x_2(t)$ and $y_2(t)$ remain 0 for all $t$. We
have taken $\lambda_1=\lambda_2=0.1$, $\alpha_1=0.6$, and $\alpha_2=0.5$. The
initial conditions are $x_1(0)=y_1(0)=1$.}
\label{f15}
\end{figure}

Figure~\ref{f15} shows that the organism does not survive if the second
antibody-antigen pair $x_2,y_2$ is not initially present. In this figure we take
$x_1(0)=y_1(0)=1$ but we we take $x_2(0)=y_2(0)=0$.

\begin{figure}[h!]
\begin{center}
\includegraphics[scale=0.38]{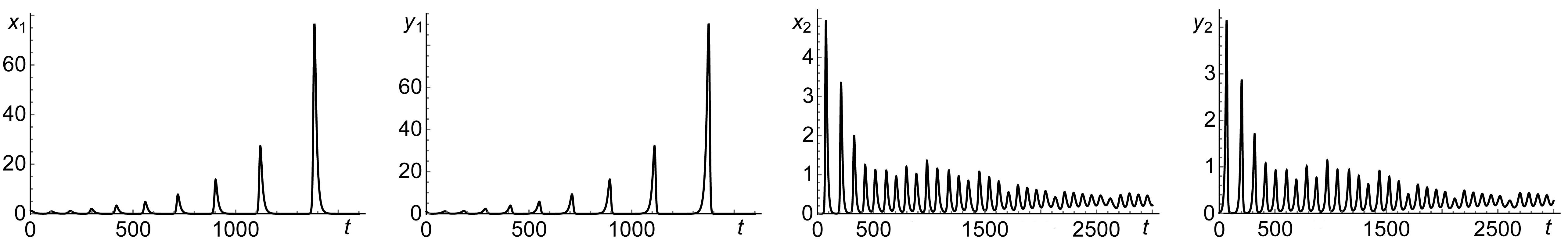}
\end{center}
\caption{Antibody-antigen competition in the broken-$\cPT$-symmetric phase in
the lower-left corner of Fig.~\ref{f14}; specifically $f=0.02$ and $g=0.01$.
The organism does not survive the antigen attack. The antigen-antibody dynamics
is described by (\ref{e11}), where $\lambda_1$, $\lambda_2$, $\alpha_1$,
$\alpha_2$, and the initial conditions are the same as in Fig.~\ref{f14}.}
\label{f16}
\end{figure}

Figure~\ref{f16} shows what happens in a broken-$\cPT$-symmetric phase when the
organism does not survive. We take $f=0.02$ and $g=0.01$, which puts us in the
lower-left corner of Fig.~\ref{f14}. The initial conditions are $x_1(0)=y_1(0)=
1$ and $x_2(0)=y_2(0)=0.01$. Note that the level of the $y_1(t)$ antigen grows
out of bounds. 

\begin{figure}[h!]
\begin{center}
\includegraphics[scale=0.38]{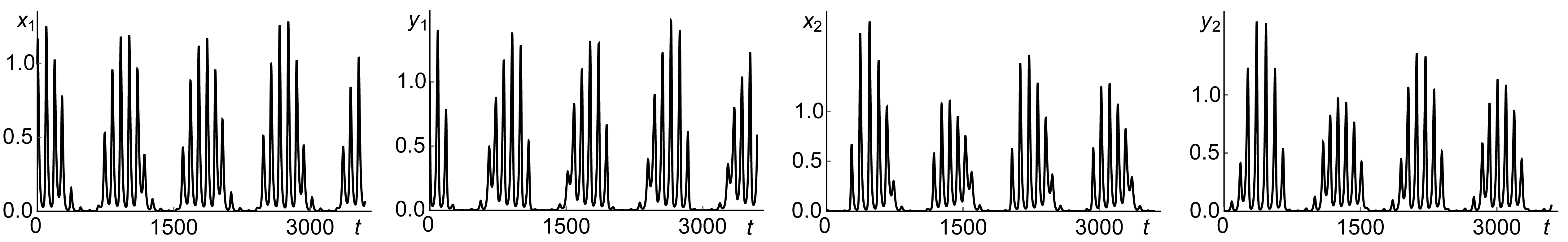}
\end{center}
\caption{An organism that survives an antigen attack. The coupling parameters
are chosen to be $f=0.32$ and $g=0.4$, which is in the unbroken-$\cPT$ phase in
the lower-left portion of Fig.~\ref{f14}. The antigen-antibody dynamics is
described by (\ref{e11}), where $\lambda_1$, $\lambda_2$, $\alpha_1$,
$\alpha_2$, and the initial conditions are the same as in Fig.~\ref{f14}. The
concentrations of antigens and antibodies behave chaotically in time.}
\label{f17}
\end{figure}

Figure~\ref{f17} shows what happens in the unbroken region in Fig.~\ref{f14}.
The organism survives but the disease becomes chaotically chronic.

\begin{figure}[h!]
\begin{center}
\includegraphics[scale=0.44]{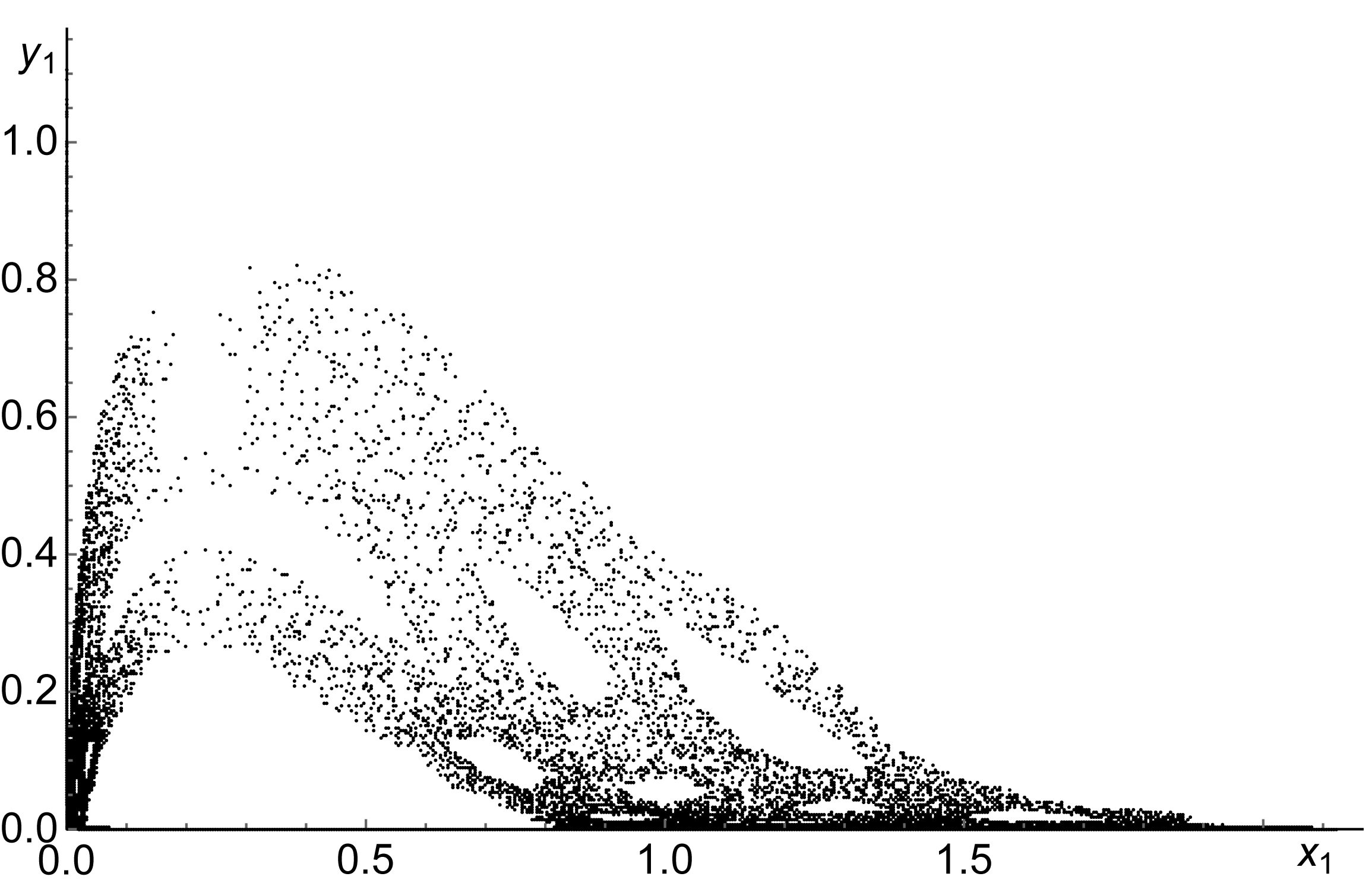}
\end{center}
\caption{An organism that survives an antigen attack. The antigen-antibody
dynamics is described by (\ref{e11}), where $\lambda_1$, $\lambda_2$,
$\alpha_1$, $\alpha_2$, and the initial conditions are the same as in
Fig.~\ref{f14}. In this plot $f=0.76$ and $g=0.80$, which places the system in
the unbroken phase in the upper-right corner of Fig.~\ref{f14}. The dynamical
behavior is chaotic and the disease becomes chronic, as implied by the
Poincar\'e map in which trajectory points are plotted in the $(x_1,y_1)$ plane
for $y_2=0.5$. The scatter of points indicates chaotic behavior. The time
interval for the plot is from $t=0$ to $t=5,000,000$.}
\label{f18}
\end{figure}

Figure~\ref{f18} demonstrates the chaotic behavior at a point in the upper-right
unbroken-$\cPT$ portion of Fig.~\ref{f14}, specifically at $f=0.76$ and $g=
0.80$. The figure shows a Poincar\'e map in the $(x_1,y_1)$ plane for $y_2=0.5$.

\begin{figure}[h!]
\begin{center}
\includegraphics[scale=0.38]{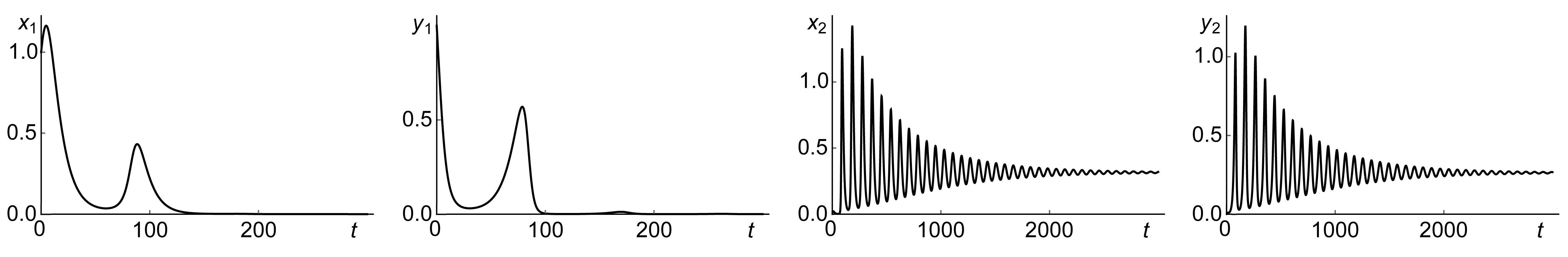}
\end{center}
\caption{An organism that survives an antigen attack. This figure shows what
happens at the point $f=0.5$ and $g=0.2$ in the broken-$\cPT$ region in the 
lower-right corner of Fig.~\ref{f14}. In this plot $\lambda_1$, $\lambda_2$,
$\alpha_1$, $\alpha_2$, and the initial conditions are the same as in
Fig.~\ref{f14}. Note that the antigen concentration $y_1(t)$ decays to zero and
the disease is cured. The concentration of antigen $y_2$ approaches a small
nonzero value as $t\to\infty$ and, as was noted in Ref.~\cite{R1}, this value is
so small that we regard it as negligible.}
\label{f19}
\end{figure}

Figure~\ref{f19} shows what happens in the broken-$\cPT$ region in the
lower-right corner of Fig.~\ref{f14} at $f=0.5$ and $g=0.2$. In this region
the antigen $y_1(t)$ completely disappears and the disease is cured.

\section{Concluding remarks}
\label{s3}
In this paper we have extended Bell's two-dimensional predator-prey model of an
immune response to a four-dimensional $\cPT$-symmetric model and have examined
the outcomes in the broken- and the unbroken-$\cPT$-symmetric phases. We have
found that in the unbroken phase the disease becomes chronic (oscillating) while
in the broken phase the host may die or be completely cured.

In Bell's model (Ref.~\cite{R1}) an oscillating regime is assumed to be a
transitory state and that either the antigen is completely eliminated at an
antigen minimum or the host dies at an antigen maximum. However, there are
many examples in which the immune system undergoes temporal oscillations
(occurring in pathogen load in populations of specific cell types, or in
concentrations of signaling molecules such as cytokines). Some well known
examples are the periodic recurrence of a malaria infection \cite{R7}, familial
Mediterranean fever \cite{R8}, or cyclic neutropenia \cite{R9}. It is not
understood whether these oscillations represent some kind of pathology or if
they are part of the normal functioning of the immune system, so they are
generally regarded as aberrations and are largely ignored. A discussion of
immune system oscillation can be found in Ref.~\cite{R10}. Additional chaotic
oscillatory diseases such as chronic salmonella, hepatitis B, herpes simplex,
and autoimmune diseases such as multiple sclerosis, Crohn's disease, and
fibrosarcoma are discussed in Ref.~\cite{R11}.

In Ref.~\cite{R1} it is not possible to completely eliminate the antigen,
that is, to make the antigen concentration go to zero. However, it is possible
to reduce the antigen concentration to a very low level, perhaps corresponding
to less than one antigen unit per host, which one can interpret as complete
elimination. However, we will see that in the $\cPT$-symmetric model (\ref{e11})
the antigen $x_1$ can actually approach 0 in the $\cPT$ broken phase.

In Ref.~\cite{R1} it is stated that the predicted oscillations of increasing
amplitude should be viewed with caution. Such oscillations are predicted to
involve successively lower antibody minima, which in reality may not occur.
However, in Ref.~\cite{R12} a modified two-dimensional predator-prey model for
the dynamics of lymphocytes and tumor cells is considered. This model seems to
reproduce all known states for a tumor. For certain parameters the system
evolves towards a state of uncontrollable tumor growth and exhibits the same
time evolution as that of $x_1$ and $y_1$ in Figs.~\ref{f15} and \ref{f16}. For
other parameters the system evolves in an oscillatory fashion towards a
controllable mass (a time-independent limit) of malignant cells. In this case
the temporal evolution is the same as that of $x_2$ and $y_2$ in Fig.~\ref{f19}.
In Ref.~\cite{R12} this state is called a {\it dormant} state. It is also worth
mentioning that in Ref.~\cite{R13} a {\it two}-dimensional dynamical system
describing the immune response to a virus is considered; this model can exhibit
periodic solutions, solutions that converge to a fixed point, and solutions
that have chaotic oscillations. Ordinarily, a two-dimensional dynamical system
cannot have chaotic trajectories but the novelty in this system is that there is
a time delay.

Finally, we acknowledge that it is not easy to select reasonable parameters if
one considers the application of Bell's model to real biological systems. In the
$\cPT$-symmetric model it is also difficult to make realistic estimates of
relevant parameters. Nevertheless, we believe that some of the qualitative
features described in this paper may also be seen in actual biological systems.

\acknowledgments
We thank M. Rucco and F. Castiglione for helpful discussions on the functioning
of the immune system. CMB thanks the DOE for partial financial support and MG
thanks the Fondazione Angelo Della Riccia for financial support.

\end{document}